 \documentclass[preprint2]{aastex}

\slugcomment{\today}

\shorttitle{Imaging Polarimetry of Mrk 3}
\shortauthors{Kishimoto et al.}

\newcommand{\chisq}{$\chi^2$}
\newcommand{\crate}{cts s$^{-1}$ pix$^{-1}$}

\begin{document}


\title{UV Imaging Polarimetry of the Seyfert 2 Galaxy Mrk 3}

\author{Makoto Kishimoto}
\affil{Physics Department, University of California, Santa Barbara, 
Santa Barbara, CA 93106}
\email{makoto@physics.ucsb.edu}

\author{Laura E. Kay}
\affil{Department of Physics and Astronomy, Barnard College, Columbia
University, New York, NY 10027}

\author{Robert Antonucci and Todd W. Hurt}
\affil{Physics Department, University of California, Santa Barbara, 
Santa Barbara, CA 93106}

\author{Ross D. Cohen}
\affil{Center for Astrophysics and Space Sciences, Code 0424, 9500
Gilman Drive, University of California at San Diego, La Jolla, CA
92093}

\and

\author{Julian H. Krolik}
\affil{Physics and Astronomy Department, Johns Hopkins University,
Baltimore, MD 21218}

\begin{abstract}

We present UV imaging polarimetry data of the Seyfert 2 galaxy Mrk 3
taken by the Hubble Space Telescope. The polarized flux is found to be
extended to $\sim$ 1 kpc from the nucleus, and the position angles of
polarization are centrosymmetric, confirming that the polarization is
caused by scattering.  We determine the location of the hidden nucleus
as the center of this centrosymmetric pattern. From the polarization
images taken in two broad bands, we have obtained the color
distribution of the polarized flux.  Some regions have blue polarized
flux, consistent with optically-thin dust scattering, but some bright
knots have a color similar to that of Seyfert 1 nucleus.  Also, the
recent Chandra X-ray observation suggests that the ratio of scattered
UV flux to scattered X-ray flux is rather similar to the intrinsic
UV/X-ray ratio in a Seyfert 1 nucleus, if the observed extended X-ray
continuum is scattered light.  While the scattered X-ray would be
essentially from electron scattering, the UV slope and UV/X-ray ratio
both being similar to Seyfert 1's would lead to two possibilities as
to the nature of the UV scatterers.  One is that the UV may also be
scattered by electrons, in which case the scattering gas is somehow
dust-free. The other is that the UV is scattered by dust grains, but
the wavelength-independent UV scattering with low efficiency indicated
by the UV slope and UV/X-ray ratio would suggest that the grains
reside in UV-opaque clouds, or the dust might be mainly composed of
large grains and lacks small-grain population.

\end{abstract}



\section{Introduction}\label{sec-intro}

Mrk 3 is one of the Seyfert 2 galaxies which have been shown to harbor
Seyfert 1 nuclei through the presence of polarized broad lines by
optical spectropolarimetry (Antonucci \& Miller 1985; Miller \&
Goodrich 1990; Tran, Miller \& Kay 1992; Tran 1995; Kay and Moran
1998; Moran et al 2000).  The polarized broad lines and featureless
continuum observed are thought to be from a hidden broad-line region
and continuum source scattered into the line of sight. Therefore, a
scattering region should exist somewhere off-nucleus in these Seyfert
2 galaxies.  High-resolution imaging polarimetry is a powerful tool to
locate this scattering region by spatially resolving the polarized
flux distribution. Also, we can directly locate the position of the
nucleus in the HST images, which is otherwise unclear since it is
obscured. The nuclear position is obtained from the distribution of
the polarization position angles, which is expected to be
centrosymmetric if the nuclear source is small enough compared to the
resolved scattering mirror (e.g. Capetti et al. 1995; Kishimoto 1999).
The exploration of the nuclear vicinity is much easier in these type 2
objects than in type 1 objects where the nucleus is too bright and
hides the surrounding regions by the wing of its point spread
function. However, this is true only if the hidden nucleus location is
robustly determined. Imaging polarimetry provides the direct method
for this determination.

Mrk 3 has an old stellar population \citep{GD01}, thus polarimetry is
highly efficient in the ultraviolet (UV) since we can avoid much of
the dilution by the unpolarized old stellar light from the host
galaxy. Thus we have implemented UV imaging polarimetry by the Hubble
Space Telescope (HST). Also, we obtained data with two different
filters to map out the color distribution of the polarized flux. This
can potentially tell us the nature of the scatterers. We describe our
observations in \S\ref{sec-obs} and the results in
\S\ref{sec-res}. The implication of the results are discussed in
\S\ref{sec-disc} and our conclusions are summarized in
\S\ref{sec-conc}. We adopt $H_0 = 65$ km sec$^{-1}$ Mpc$^{-1}$
throughout this paper. Mrk 3 is at $z=0.0140$, so the distance is 65
Mpc and $1''$ corresponds to $\sim$ 310 pc.

\section{Observation and data reduction}\label{sec-obs}

Mrk 3 was observed on December 10, 1998 by the Faint Object Camera
(FOC) onboard the HST. The filters F275W ($\lambda \sim 2800$\AA) and
F342W ($\lambda \sim 3400$\AA) were used with three polarizers
POL0/POL60/POL120. The F342W images were taken in the normal $512
\times 512$ pixel mode, where the pixel size is $0.''014 \times
0.''014$, giving a field of view of $\sim 7'' \times 7''$.  The
F275W images were taken in the zoomed mode, resulting in a larger
field of view with $1024 \times 512$ pixels, $\sim 14'' \times 7''$.
The data are summarized in Table \ref{tab-data}. All of these were
taken after the installation of COSTAR, i.e., after the refurbishment
of the HST.

The FOC suffers from nonlinearity when count rates are large, and the
nonlinearity depends on the distribution of the illumination over the
detector. It exhibits nonlinearity at the 10\% level when the count
rate is 0.08 and 0.15 \crate\ for the $512 \times 1024$ and $512
\times 512$ format, respectively, for uniform illumination. For a
point-source, the 10\% nonlinearity occurs for a peak count rate of
0.5 and 1.0 \crate, respectively. The bright regions in the images of
Mrk 3 are clumpy, so the detector behavior is expected to be somewhere
between these two illumination cases.  The maximum count rate in our
images was 0.03 \crate\ with the F275W filter ($512 \times 1024$
format) and 0.11 \crate\ with the F342W filter ($512 \times 512$
format).  Therefore, the nonlinearity in these images is expected to
be very small.  Its effect on the $P$ measurement should be
concentrated only at the positions of the few brightest clumps. Even
in these clumps, a conservative upper limit of the effect is found to
be about 1/20 and 1/10 of the measured $P$ in the F275W and F342W
images, respectively, utilizing the nonlinearity relation for uniform
illumination (Nota et al. 1996; Jedrzejewski 1992).

The data were processed in the standard manner to correct for
geometric distortion and flat-field response. The reseau marks were
removed using neighboring pixels. The three images with three
polarizers are known to be slightly shifted relative to one
another. In our FOC images of Mrk 3, there are no point sources we can
use for the image registration. Therefore, we used the image shift
calibration results from Hodge(1993,1995) which are accurate to $\pm
0.3$ pixel. The background was subtracted using the outer region of
the images.  Before combining the three images, each image was
scaled in order to allow for the different transmittances of the three
polarizers.  We have estimated the effective transmittances of each
polarizer plus filter, using large-aperture spectra from the HST/FOS
data (Cohen et al. 2001) and the ground-based data (Kay 1994).  Then
the three images through the three polarizers were scaled accordingly,
and combined to produce the Stokes $I, Q, U$ images. The polarized
flux and polarization are calculated as $\sqrt{Q^2 + U^2}$ and
$\sqrt{Q^2 + U^2} /I$, respectively, and debiased following Simmons
and Stewart (1985).

Using the same large-aperture spectra, we also estimated the narrow
emission line contamination in these filters to be $\sim 30$\% for
both of the filters. The F275W filter is primarily affected by the
MgII $\lambda$2800\AA\ line and the F342W filter by [OII]
$\lambda$3727\AA, [NeV]$\lambda$3426, and [NeV]$\lambda$3346 lines.
This affects the absolute flux measurement of the continuum
accordingly. For small spatial bins, the line contamination could be
different from this estimation.  In terms of $P$ measurement, the line
contamination simply results in diluting $P$, since the narrow lines
are not strongly polarized [essentially only containing the foreground
interstellar polarization in our Galaxy (see below; Schmidt \& Miller
1985, Goodrich 1992), although they could be slightly polarized
intrinsically (Tran 1995)].  The contamination by unpolarized lines
essentially will not affect the $Q$ and $U$ measurements (though the
small spatial scale variation of the effective transmittance would
slightly affect the $Q$ and $U$ measurements: the resulting
uncertainty in $P$ is estimated to be less than $\sim$ 1\%).

However, our two filters are on the so-called 3000\AA\ bump, which
consists of broad FeII lines and Balmer continuum plus high-order
Balmer emission lines. The relative polarized flux color measurement
between different locations would not be affected by these
contaminations, since the incident spectrum from the hidden nucleus is
the same, but the absolute color measurement will be slightly
affected. We will discuss this in \S\ref{sec-disc}.

The optical interstellar polarization in our Galaxy toward Mrk 3 has
been estimated to be $\sim$ 1.2\% at PA = 132\degr, from the
polarization of the nearby foreground stars (Schmidt \& Miller 1985).
We have corrected our polarization maps for this foreground
polarization, using the Serkowski curve (Serkowski, Mathewson, \& Ford
1975) with the parameters adopted by Tran (1995) which are based on
the measurement of Schmidt \& Miller.

The images through the F275W and F342W filters are also shifted to
each other.  We have registered them by taking cross-correlation of
the central $\sim 4''$ region of the $I$ images produced above, since
both images are dominated by the same clumpy structure in the central
region. The uncertainty in this registration is estimated be $\pm 0.5$
pixel.

There are various error sources in the FOC imaging polarimetry. These
are described in detail in Kishimoto (1999). Briefly, there are four
major sources : (1) statistical error (2) uncertainty in the image
registrations of the three polarizer images (3) uncertainty in the
polarizer axes direction (4) uncertainty in the relative intensities
through each polarizer, mainly from the differences in the shape of
the point spread function (PSF) through each polarizer. The source (4)
becomes a major error source when the synthetic aperture or binning
size for measuring polarization is small. For many cases, the error
sources (1) and (4) are larger than others. For (1), Poisson noise is
assumed. The source (4) depends on the binning size in the
polarization calculations. In this paper, we mainly use a 10 pixel bin
($\sim 0.''14$) and a 40 pixel bin ($\sim 0.''57$), and adopt
uncertainties of 5\% and 2\% for the error source (4) in these bins,
respectively (the former is the same value as in Kishimoto 1999, and
the latter is an extrapolation from the value for the 10 pixel bin and
the 20 pixel bin). In addition to these error sources, we also added
in quadrature 5\% of the background subtraction amount as an error in
the counts in each binned image, in order to evaluate the polarization
measurement uncertainty in the diffuse outer regions.

\section{Results}\label{sec-res}

\subsection{Extended scattering region}

\begin{figure}
\plotone{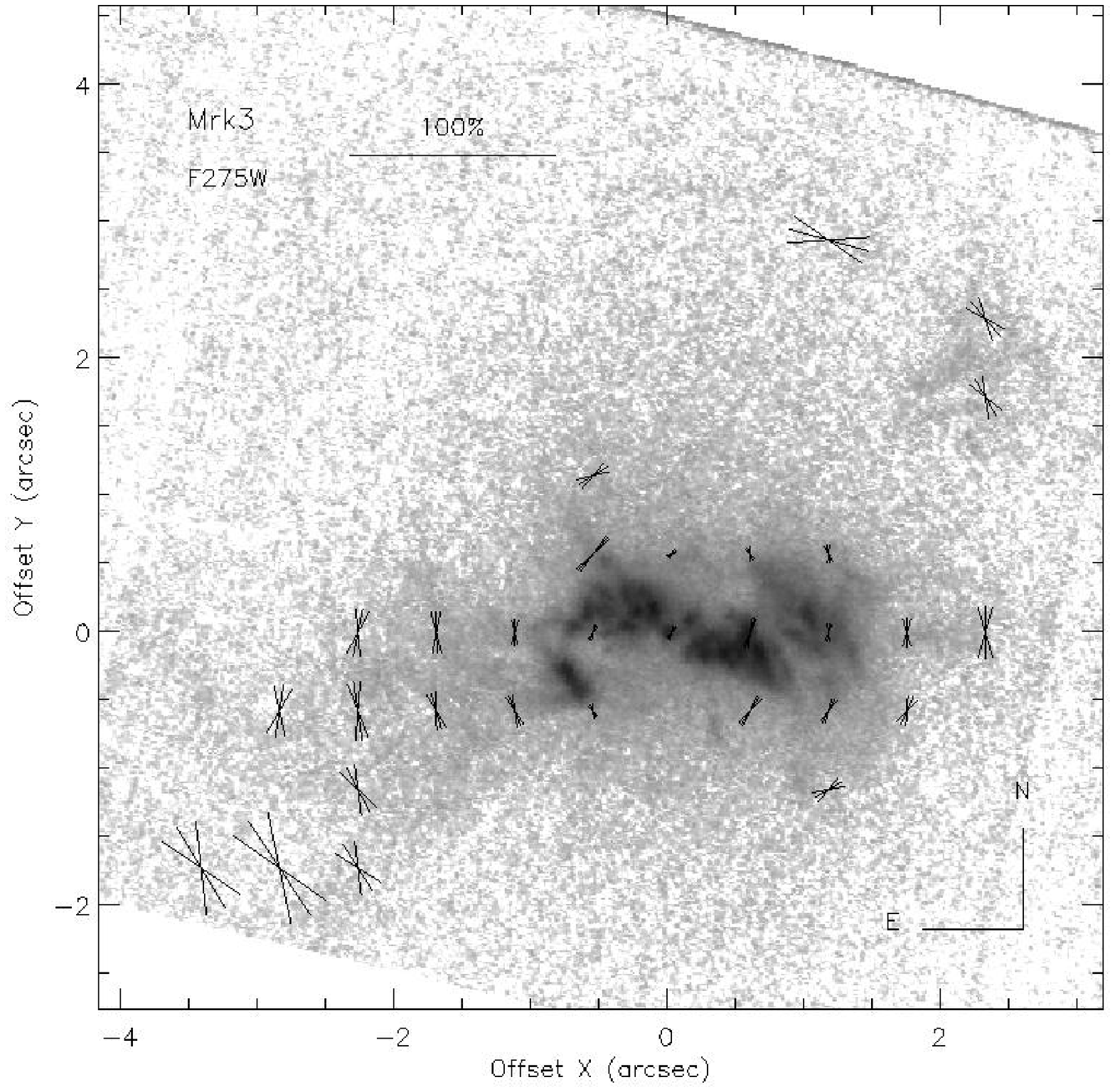}
\figcaption{Polarization map of Mrk 3 with F275W. The polarizations
are calculated in 40 pixel ($0.''57$) bins. The directions of the
lines at each position represent $\theta_{\rm PA}$ (direction of
E-vector) and $\theta_{\rm PA} \pm \sigma_{\theta}$. The regions with
statistical S/N in $P$ larger than 5 are shown, but the errors shown
are the total sum of the various error sources in quadrature,
including systematic errors (see \S\ref{sec-obs}).  The lengths of the
lines are proportional to the polarization degree, where $1.''5$
length corresponds to 100\%. The underlying grayscale image is the
total intensity with the F275W filter in log scale. North is up, east
is to the left. The origin of the coordinates is taken at the
intensity peak of the cloud closest to the nucleus in this and all
subsequent figures (see Fig.\ref{fig_mrk3_nucpos}).
\label{fig_mrk3_PA_L}}
\end{figure}

Figure \ref{fig_mrk3_PA_L} shows the polarization map of Mrk 3 with
the F275W filter. The polarizations are calculated in 40 pixel
($0.''57$) bins, and the regions with statistical S/N in $P$ larger
than 5 are shown. The position angle errors indicated in the figure
are the total sum in quadrature of all the error sources described in
the previous section.  The position angle (PA) distribution is
centrosymmetric, supporting the idea that the polarization is caused
by the scattering of the radiation from a compact nuclear source.  The
lengths of the lines at each position in the figure indicate the
degrees of polarization. We note that in the outermost regions where
uncertainty from the background subtraction dominates, the
polarization degree is highly uncertain accordingly. However, at least
the PA directions are consistent with the ones in the inner regions,
suggesting that we have detected the polarization even out to $\sim
3''$ from the center.

\begin{figure}
\plotone{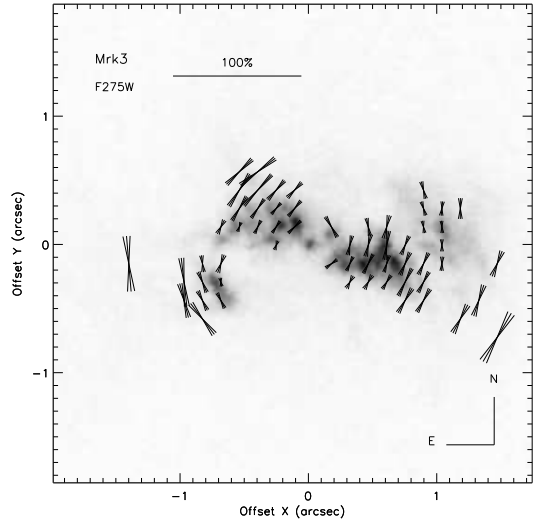}
\figcaption{The central $\sim 4''$ region of the polarization map of
Mrk 3 in the F275W filter with 10 pixel ($0.''14$) bins. The notations
for the polarization are the same as Fig.\ref{fig_mrk3_PA_L}, and
$1''$ length corresponds to 100\% as indicated. The regions with
statistical S/N in $P$ larger than 5 are shown, but the errors
indicated include other error sources as in
Fig.\ref{fig_mrk3_PA_L}. The grayscale image is the total intensity
with the F275W filter in linear scale.
\label{fig_mrk3_P_S275}}
\end{figure}

\begin{figure}
\plotone{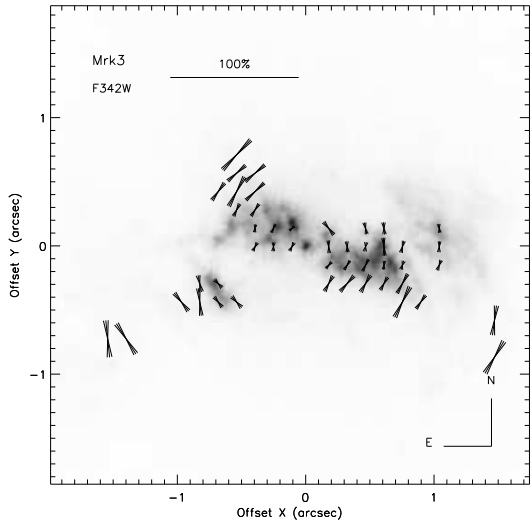}
\figcaption{The central $\sim 4''$ region of the polarization map of
Mrk 3 in the F342W filter with 10 pixel ($0.''14$) bins. The notations
are the same as Fig.\ref{fig_mrk3_P_S275}. The grayscale image is the
total intensity with the F342W filter in linear
scale. \label{fig_mrk3_P_S342}}
\end{figure}

We also show the polarization map on a smaller scale in Figures
\ref{fig_mrk3_P_S275} and \ref{fig_mrk3_P_S342}. The polarizations are
calculated in 10 pixel bins ($0.''14$), and only the regions with
statistical S/N in $P$ larger than 5 are shown.  The lengths of the
lines are proportional to the polarization degrees also in these
figures. The central $\sim 1''$ radius region is very clumpy, and the
PA pattern is centrosymmetric down to this central region, where $P$
is $10-20$\% and $5-15$\% level in the F275W and F342W filter,
respectively (though note the dilution of $P$ by the narrow line
contamination; see \S\ref{sec-obs}).

\subsection{Centrosymmetric pattern and the location of hidden
nucleus}

\begin{figure}
\plotone{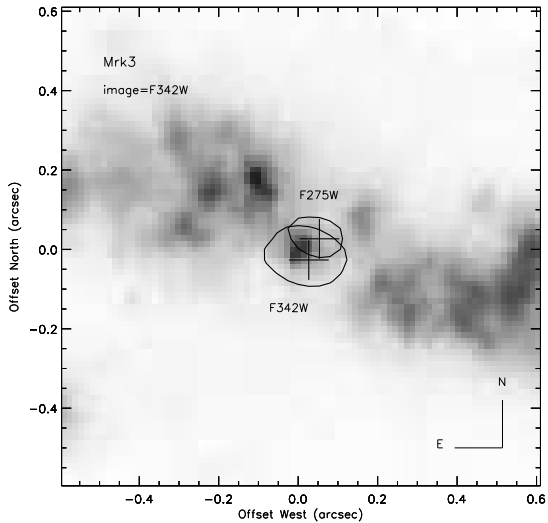}
\figcaption{The location of the hidden nucleus in Mrk 3. The minimum
\chisq\ point (plus sign) and the error circle of 99\% confidence
level are indicated for each of the F275W and F342W filter data. The
pair at the NW side is for the F275W data, and the other pair at the
SE side is for the F342W data.  The grayscale image is the total
intensity with the F342W filter in linear
scale. \label{fig_mrk3_nucpos}}
\end{figure}

From the observed centrosymmetric pattern, we can robustly determine
the position of the hidden nucleus as a symmetric center of this
pattern. Specifically, assuming a centrosymmetric model of the PA
distribution, a least square fit is implemented using the PA
measurements with given errors at each point of the image. The only
parameters are the position of the symmetric center. The most probable
location of the nucleus is determined as the point of the minimum
\chisq, with an error circle defined by a certain confidence
level. The method is described in detail in Kishimoto (1999).  We have
implemented this fit to the data shown in Figure \ref{fig_mrk3_P_S275}
and \ref{fig_mrk3_P_S342}.  The reduced \chisq\ was found to be
$66.3/56=1.18$ and $48.3/45=1.07$ for the F275W and F342W data,
respectively.  The minimum \chisq\ point is shown as a plus sign in
Figure \ref{fig_mrk3_nucpos} together with the error circle of 99\%
confidence level, for each of the filters.  The two results are
slightly different, though they are not inconsistent with each other.

The central $\sim 2''$ of the UV image of Mrk 3 consists of several
resolved clouds. There is no unresolved, overwhelmingly bright point
source in the image, as expected from the unified model. However, the
error circle for the nuclear location includes (or marginally
includes) a rather bright cloud. The nucleus is thought to be hidden
from direct view in this wavelength region, so we would not expect the
radiation of this cloud to be the direct nuclear light.  The FOC data
rather prefer that the nucleus resides at the western side (SW or NW)
of the cloud, but the data do not strongly constrain which side the
nucleus is on, and do not even exclude the possibility that the
nucleus is within the cloud.  The size of this cloud is about $0.''06$
($\sim 20$pc) in FWHM in both of the F275W and F342W images, where the
diffraction limit of the HST at this wavelength is about $0.''03$. 

The total flux color of this cloud is very red : the synthetic
photometry with a circular aperture of $0.''15$ diameter gives
$F_{\lambda}$(2800\AA)/$F_{\lambda}$(3400\AA) = $0.55\pm0.02$, which
corresponds to $\alpha = -5.0 \pm 0.2$ where $F_{\nu} \propto
\nu^{\alpha}$. The measured flux is $F_{\lambda}$(2800\AA) =
$(3.6\pm0.1) \times 10^{-17}$ erg cm$^{-2}$ s$^{-1}$ \AA$^{-1}$ and
$F_{\lambda}$(3400\AA) = $(6.5\pm0.2) \times 10^{-17}$ erg cm$^{-2}$
s$^{-1}$ \AA$^{-1}$ (but note the emission line contamination; see
previous section). These are after the correction for the Galactic
reddening, using the reddening curve of Cardelli, Clayton, \& Mathis
(1989) with $E_{B-V} = 0.188$ (NED; Schlegel et al. 1998).

The calculated polarization for this cloud is not much larger than the
uncertainty of the FOC polarization measurement with small apertures.
The synthetic circular aperture polarimetry with $0.''15$ diameter for
this blob, after the interstellar polarization correction (see
\S\ref{sec-obs}), gives $P = 3.6 \pm 4.4$\% (statistical error 1.4\%)
and $\theta_{\rm PA} = 2 \pm 37\degr$ (statistical error $11\degr$) at
F275W, $P = 5.1 \pm 4.3$\% (statistical error 1.1\%) and $\theta_{\rm
PA} = 6 \pm 25\degr$ (statistical error $6\degr$) at F342W.
Therefore, the polarization detection is marginal, but the blob could
be polarized with the $4-5$\% level, since our error estimation would
be conservative, as discussed in Kishimoto (1999).

\subsection{The color map of polarized flux}

\begin{figure}
\plotone{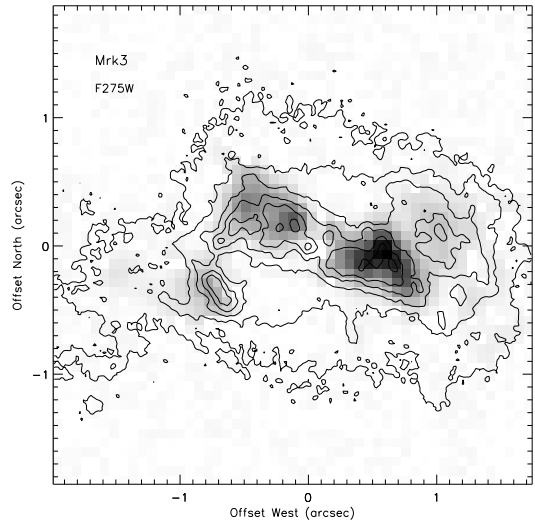}
\figcaption{The polarized flux distribution with F275W filter. The
three images through three polarizers were convolved with a Gaussian
of FWHM 10 pixel ($0.''14$), and polarized flux is calculated with a 5
pixel bin, and shown in linear grayscale. The contours are the $I$
image with the F342W filter (not the F275W filter), convolved with a
Gaussian of FWHM 2 pixel. \label{fig_mrk3_f275_pf}}
\end{figure}

\begin{figure}
\plotone{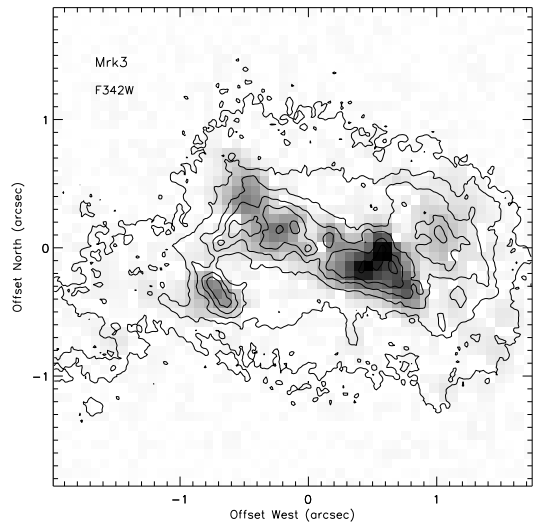}
\figcaption{The same as Fig.\ref{fig_mrk3_f275_pf}, but for the F342W
filter. The grayscale is linear, with a peak at the same pixel as in
Fig.\ref{fig_mrk3_f275_pf}.
\label{fig_mrk3_f342_pf}}
\end{figure}

Figures \ref{fig_mrk3_f275_pf} and \ref{fig_mrk3_f342_pf} show the
polarized flux distribution through the F275W and F342W filter,
respectively. The three polarizer images for each filter were smoothed
by a Gaussian with a FWHM of 10 pixels ($\sim 0.''14$), and polarized
flux was calculated with a 5 pixel bin.  The polarized flux has been
debiased following Simmons \& Stewart (1985); namely corrected by a
factor of $(1 - (\sigma_P/P_{\rm obs})^2)^{1/2}$ ($\sigma_P$ is a
statistical error, $P_{\rm obs}$ is an observed polarization). The
regions with $P_{\rm obs}/\sigma_P < 1$ are masked out.  The contours
of the $I$ image with the F342W filter are drawn in both of the
figures, to make the comparison of the two polarized flux distribution
easier. The grayscale is linear in both images, with a peak at the
same pixel [$\sim (0.''6, -0.''1)$]. The peak polarized flux in the
F342W filter is lower by a factor of 0.75 than that in the F275W
filter (after the Galactic reddening correction with $E_{B-V} =
0.188$). Most of the polarized flux (roughly $\sim 70$\%) is coming
from the region within $\sim 1''$ ($\sim$ 300 pc) from the nucleus,
and the polarized flux is greater on the west side than the east side.

\begin{figure}
\plotone{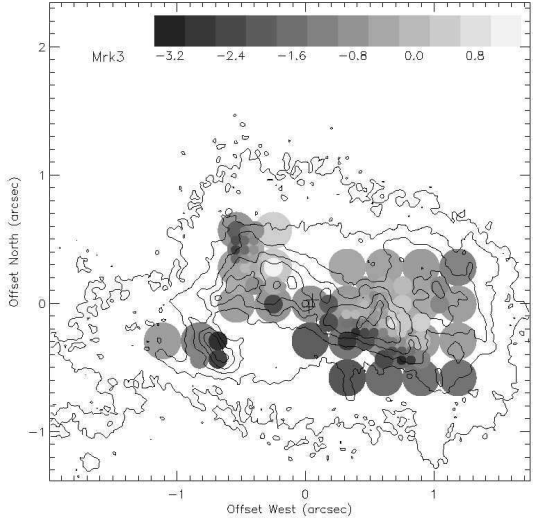}
\figcaption{The color map of polarized flux in Mrk 3. The ratio of the
polarized flux in F275W to that in F342W has been converted to the
spectral index $\alpha'$ where $F_{\nu} \propto \nu^{\alpha'}$ (this
color is slightly different from the color of the true polarized
continuum; see text). The palette shows the correspondence between the
colors and $\alpha'$ values. Each polarizer image was smoothed with
FWHM 40, 20, 10 pixel Gaussian, and three images with different
polarizers were combined to calculate the polarized flux with 20, 10,
5 pixel bins, respectively.  The small bin measurements are simply
superposed on the larger bin measurements. Note that the binned pixels
are square, but the colors are illustrated by circles in order to make
the distinctions between small and large bins clear. The polarization
centers in Fig.\ref{fig_mrk3_nucpos} for both of the F275W and F342W
data are shown as plus signs. \label{fig_mrk3_pfcolor}}
\end{figure}

The polarized flux distribution is different in these two filters.
From the images with these two filters, we can derive the color
distributions of the polarized flux, which is shown in Figure
\ref{fig_mrk3_pfcolor}. The ratio of the polarized flux in the F275W
filter to that in the F342W filter has been converted to the spectral
index $\alpha$ ($F_{\nu} \propto \nu^{\alpha}$).  The spectral index
has been corrected for the Galactic reddening of $E_{B-V} = 0.188$
(NED; Schlegel et al. 1998). The resulting index range shown in the
figure is from $-3.1$ to $+1.2$. This color map is a composite of
three different bins with three different smoothing. We have
convolved three polarizer images for each filter with a Gaussian of
FWHM 40, 20, 10 pixel and generated the color map with 20, 10, 5 pixel
bins, respectively, and stacked them into one plot.  For each bin
case, the regions with the formal 1-$\sigma$ uncertainty (calculated
using the smoothed image counts) in the spectral index smaller than 2
are shown, but the actual uncertainty for the regions shown is
estimated to be less than $\sim 1$ by binning the images with the
smoothing FWHM size.  For the error calculation, we have neglected the
error source (4) described in \S\ref{sec-obs}, because of the rather
heavy smoothing.  The caveat is that the color at the regions with
large intensity gradient should be taken with caution, since it would
have influence from the spatial blurring of the bright region.  We
also made the color map without any smoothing and obtained a
consistent map, though the features are much clearer in the map from
smoothed images. Therefore we only show the latter.

\begin{figure}
\plotone{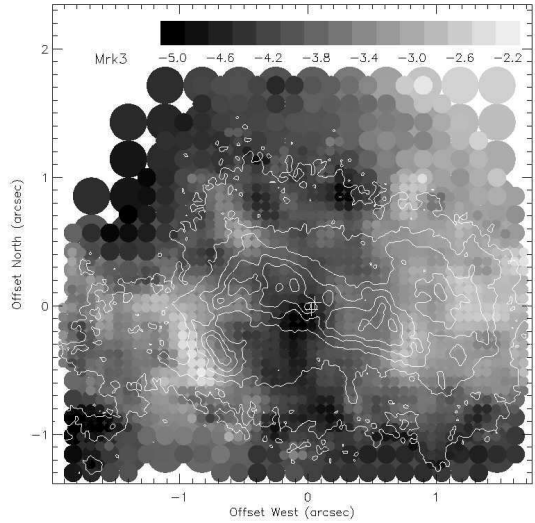}
\figcaption{The same as Fig.\ref{fig_mrk3_pfcolor} but for total flux.
Note that the color range is redder than in
Fig.\ref{fig_mrk3_pfcolor}.  The color would be slightly different
from the true continuum, due to the line contamination in both of the
two filters.
\label{fig_mrk3_tfcolor}}
\end{figure}

We also generated the total flux color map, which is shown in Figure
\ref{fig_mrk3_tfcolor}.  The color has been corrected for the Galactic
reddening. The same procedure as for the polarized flux color map was
taken, except the threshold for the formal 1-$\sigma$ uncertainty in
the index which was set to 1. In the central $1''$ radius region, the
actual uncertainty is estimated to be less than 0.3.  The spectral
index range is found to be $-5.0 \sim -2.2$, significantly redder than
the polarized flux. The color is very red at around the location of
the hidden nucleus ($\sim -5.0$; see the previous subsection), and
this red color is extended to the south. This is probably due to an
enhanced extinction. The red color seems to be also extended along the
north-south direction. On the other hand, the color tends to be bluer
at the regions adjacent to some bright clouds, on the opposite side of
the direction to the nucleus : especially the east side of the
southeastern blob and the west side of the western bright blob.
However, the interpretation of the total flux color in general is
uncertain : while the overall red color could be partly due to the
contribution from old stellar population in the host galaxy, the
small-spatial-scale color variation could be caused by the narrow-line
contamination in our two filters (see \S\ref{sec-obs}). Therefore we
will not attempt to interpret the total flux color map in this
paper. We will discuss the polarized flux color map in
\S\ref{sec-disc}.

\section{Discussion}\label{sec-disc}

\subsection{The nuclear location and extended scattered flux}

Mrk 3 shows a linear jet-like structure in the radio at PA 84\degr,
consisting of several knots. One central knot remains unresolved even
with a high spatial resolution of $0.''035$ at 5GHz, and this is
thought to be the nuclear location (Kukula et al. 1993). Therefore, to
register the radio map on to the HST image, which typically has
uncertainties of $\sim 0.''5$, this radio core should fall within the
error circle of the polarization center as determined above. The
registration adopted by Capetti et al. (1996) is consistent with this
(while the registration in Capetti et al. 1995 is slightly different).

The spatial distribution of the polarized flux provides a constraint
on the opening angle of the nuclear anisotropic radiation. From
Figures \ref{fig_mrk3_P_S275} and \ref{fig_mrk3_P_S342}, the projected
full opening angle is $\sim 110\degr$, from PA $\sim$ 25\degr\ to PA
$\sim$ 135\degr. Obviously, the region illuminated by the nucleus is
not filled with bright narrow-line emitting gas. The morphology of the
narrow-line emitting cloud distribution is not determined in detail by
the anisotropic radiation pattern. Instead, it is suggested to be
closely related to the radio jet (Capetti et al. 1995, 1996, 1999).
We note that the larger scale biconical radiation morphology (Pogge \&
De Robertis 1993), extended out to $5''$ from the nucleus, is at PA
114\degr\ with a full opening angle of $\sim$ 74\degr, and thus
slightly shifted from the small scale radiation structure.

\subsection{Polarized flux color and a Seyfert 1 color}

The color map of the polarized flux can potentially tell us the nature
of the scatterers. The obtained color map (Fig.\ref{fig_mrk3_pfcolor})
seems to suggest that the scattering mechanism is rather non-uniform
and complicated.  As described in \S\ref{sec-obs}, our two filters are
on the 3000\AA\ bump: the polarized flux through the F275W filter
includes the broad FeII and MgII lines, and the F342W filter contains
the Balmer continuum. The relative color measurement at different
spatial locations is not affected by these contaminations, since the
incident spectrum for the scattering region is the same --- the one
from the hidden nucleus. The interpretation of the absolute value,
however, needs a reference value for a Seyfert 1 nucleus.

For this purpose, we have collected archival HST/FOS spectra of 14
Seyfert 1 galaxies ($M_B > -23$) in this wavelength range, taken with
small apertures ($0.''86$ diameter, and $0.''3$ for a nearest
one). The color between the F275W and F342W filters was derived by
synthetic photometry, and converted to the spectral index $\alpha$
($F_{\nu} \propto \nu^{\alpha}$). For a few objects for which
forbidden lines such as [OII]$\lambda$3727 and [NeV]$\lambda$3426
seemed significant, we have removed these lines, but their effect was
found to be only less than $\sim 2$\%.  Combined effect of FeII and
MgII lines at F275W would be slightly larger than the effect of the
Balmer continuum at F342W (contamination by MgII alone is about 5\%),
so this color may be slightly bluer than the true continuum.  We
denote the obtained spectral index as $\alpha'$.

The obtained $\alpha'$ values range from $-1.4$ to $+0.1$ where the UV
luminosity $\nu L_{\nu}$ at 2800\AA\ is in the range of $10^{43} \sim
3 \times 10^{45}$ erg s$^{-1}$, but there is a tendency for
low-luminosity ones to have redder colors, similar to the correlation
found in Mushotzky \& Wandel (1989). The host galaxy contamination
would make the color redder, but in our case, this is perhaps avoided
by the small apertures used.  Mrk 3 is estimated to have, if
unobscured, $\nu L_{\nu}$ at 2800\AA\ around $1\times10^{44}$ erg
s$^{-1}$ from the direct hard X-ray component, assuming a fiducial
UV/X-ray ratio of 10 for a Seyfert 1 galaxy (see below). We also get a
consistent, approximate upper limit for this UV $\nu L_{\nu}$ to be
$\sim 3 \times 10^{44}$ erg s$^{-1}$ from the IRAS measurement at 25
\micron, assuming the ratio of $\nu F_{\nu}$ at 2800\AA\ to 25
\micron\ to be $\sim$ 2 (Sanders et al. 1989). Therefore, as a
reference value of $\alpha'$, we take the average for the five objects
with $10^{43}$ erg s$^{-1}$ $< \nu L_{\nu} < 3\times10^{44}$ erg
s$^{-1}$ , which is $-0.9$.

\subsection{Color map interpretation and constraints from the
Chandra observation}

One obvious feature in our polarized flux color map is that the color
is significantly redder than the Seyfert 1 color in the southern edge
of the bright knotty regions.  This red color is probably due to
foreground extinction in this southern edge region, where a dust lane
feature is seen in the HST optical continuum image (Capetti et
al. 1996).  In contrast, the polarized flux color seems to be bluened
in some regions : the west region of the west bright knots ($\alpha' =
+0.9$ at $\sim 0.''7$ west of the nucleus), and the north region of
the east bright knots ($\alpha' = +1.2$ at $\sim 0.''4$ north-east of
the nucleus). This blue color suggests that the polarized flux in
these regions could be from optically-thin dust scattering, though the
spatial extent of the blue regions at the edges of the bright regions
should be taken with caution, because of the spatial blurring by
smoothing.  The spectral index change expected for Galactic type dust
scattering is $1.0-1.5$ (see e.g. Kishimoto et al. 2001 and references
therein), and this seems to be roughly consistent with the blue color
seen in Figure \ref{fig_mrk3_pfcolor}, considering the typical Seyfert
1 color quoted above.  However, scattered light in the whole nuclear
region is not dominated by this blue radiation. In the central
brightest regions, the color of some bright knots, e.g. the west
brightest knots ($\alpha'=-0.8$), is similar to the Seyfert 1 color,
which would suggest gray scattering. Synthetic photometry with a
$0.''8 \times 0.''6$ aperture on the west brightest knotty region
gives $\alpha' = -0.7 \pm 0.5$, while the same photometry on the east
knotty region yields a little bluer color, $\alpha' = -0.1 \pm
0.6$. The central $2.''5 \times 1.''4$ region, which includes the
whole bright region, has $\alpha' = -0.5 \pm 1.0$ (the error is
dominated by the background subtraction uncertainty).

One interpretation for this polarized flux color range would be that
the scatterers are dominated by dust grains and the color distribution
is all from reddening. The reddening could be foreground or
intermingled within the scattering region itself.  While the red color
found in the southern edge region would be from a foreground
reddening, the whole reddening would not be dominated by a foreground
reddening, since the FOS UV polarized flux (taken with $4.''3 \times
1.''4$ aperture, 2200-3200\AA; Cohen et al. 2001) does not show an
exponential decrease toward shorter wavelengths. On the other hand,
scattering intermingled with reddening tends to make the
scattering wavelength dependence rather gray, without an exponential
decrease.  Independent spatially-resolved reddening maps, such as an
H$\alpha$/H$\beta$ ratio map, would be able to test this reddening
explanation, though such a map is not yet available.

The recent Chandra X-ray observation of Mrk 3 (Sako et al. 2000) found
a spatially extended soft X-ray continuum component. Its size is $\sim
2''$ ($\sim 600$pc), and its spatial extent seems to coincide with
that of the dominant polarized flux in our UV image. Sako et al. find
the flux of this extended continuum to be $4.3 \times 10^{-13}$ erg
cm$^{-2}$ s$^{-1}$ (converted from the quoted luminosity with their
adopted $H_0$ value of 65 km s$^{-1}$ Mpc$^{-1}$), integrated between
1 and 1000 Ryd assuming the photon index of $\Gamma = 1.8$. This
converts to $\nu F_{\nu} = 6.7 \times 10^{-14} (\nu/\nu_{\rm 1
keV})^{0.2}$ erg cm$^{-2}$ s$^{-1}$.  If this is considered to be
scattered light, as suggested by Sako et al., we can combine it with
the UV scattered flux to see how this scattered flux shape between UV
and X-ray matches that of a typical Seyfert 1 spectrum.  Note that the
scattered X-ray is considered to be from electron scattering, since
X-ray scattering by dust is so forwardly concentrated (scattering
angle is very small) that dust-scattered X-ray will not get into our
line of sight in the case of type 2 objects such as Seyfert 2
galaxies.  From the F275W filter image, we deduce a total flux of $\nu
F_{\nu} = 7.1 \times 10^{-12}$ erg cm$^{-2}$ s$^{-1}$, and a polarized
flux $\nu F_{\nu} = 6.5 \times 10^{-13}$ erg cm$^{-2}$ s$^{-1}$ at
2800\AA\ (within the central $2.''5 \times 1.''4$ region). The UV
scattered flux should be somewhere between these two. Therefore, the
ratio of the UV to X-ray scattered flux in $\nu F_{\nu}$ is $\sim 10 -
100$, whereas the typical ratio seen in radio-quiet quasars with
luminosity less than $10^{12} L_{\odot}$ is roughly $\sim$ 10 (ratio
of $\nu F_{\nu}$ at $\sim 3000$\AA\ to $\sim$ 1keV; Sanders et
al. 1989).

This ratio is apparently consistent with electron scattering, since
the Thomson scattering cross section $\sigma_T$ does not change with
wavelength and the spectral shape will not be changed in this case,
although the data do allow the possibility that the UV/X-ray scattered
light ratio is a little larger than in most Seyfert 1s.  On the other
hand, in order to explain this ratio by having Galactic-type dust
scattering in the UV, the UV scattered light needs to have been
suppressed somehow considerably. This is because the UV scattering
cross section of the Galactic dust grains per H atom is much larger
than $\sigma_T$ (roughly $\sim$ 500 $\sigma_T$ in our UV range, though
of course it depends on the sight lines), while in X-ray, the
scattering cross section is essentially $\sigma_T$.

The suppression could come from the possible opaqueness of the
scattering region.  The X-ray observation suggests a rather high
column density for the scattering region. In Mrk 3, the direct light
from the nucleus dominates in the hard X-ray, which gives us the
intrinsic nuclear X-ray luminosity. Comparing this direct flux ($\nu
F_{\nu} \sim 2 \times 10^{-11}$ erg cm$^{-2}$ s$^{-1}$ at 1 keV,
absorption corrected) with the scattered X-ray luminosity quoted
above, Sako et al. estimate the scattering optical depth in X-rays to
be $\sim 0.01$ (assuming conical scattering region of half opening
angle 50\degr; this corresponds to the average column density of
$1.5\times10^{22}$ cm$^{-2}$ using $\sigma_T$: note the typo in Sako
et al. 2000).  This would mean that the average UV extinction optical
thickness is $\sim$ 10 for Galactic-type dust of the normal
dust-to-gas ratio (scattering optical thickness of $\sim 5$ with
albedo $\sim 0.5$).  In the HST images, the scattering region is not
filled uniformly, but instead it is clumpy and the covering factor
could be small, so the column density through each cloud would be even
larger than this average column density.

Once the clouds become opaque at the UV, the amount of the UV
scattered light will not essentially depend on the column density of
each cloud, but will approximately saturate at the amount for the
extinction optical thickness around unity. Then the scattered light
amount will essentially be proportional to the covering factor.
Therefore, given the same average column density, if we consider a
case of smaller covering factor (e.g. the case that the resolved
clouds consist of smaller unresolved clouds) with each cloud having
larger column density, the scattered light will be reduced in the UV,
but not in the X-ray, as long as the clouds do not get also opaque in
the X-ray.

Thus, scattering by clumpy UV-opaque clouds, as well as electron
scattering, would be consistent with the observed UV/X-ray scattered
flux ratio.  However, the extreme case of a very small covering
factor, with the clouds being X-ray opaque, i.e. Compton-thick
(covering factor of 1\% makes the column density already of order
$10^{24}$ cm$^{-2}$), seems to be ruled out, since in this case the
X-ray scattered continuum from such clumps would be flatter (photon
index $\Gamma \sim 0$ or bluer at $\sim 1-10$ keV for matter that is
not too highly ionized; e.g. Miller, Goodrich, \& Mathews 1991; Ross
\& Fabian 1993) than observed ($\Gamma$ for the extended scattered
component is larger than 1.2; M. Sako, private communication 2001; the
CCD spectrum from the zeroth order X-ray image is also consistent with
this).  On the other hand, electron scattering might give a simple
explanation for the UV to X-ray scattered flux ratio and the gray
color in the bright knots.  If this is the case, the gas in these
regions should have lost its dust content almost completely.  This
could imply that this electron scattering gas came as a wind from the
nucleus, where the dust equilibrium temperature is above the
sublimation temperature, though it might be odd that the gas from a
wind apparently resides in knotty regions.

\subsection{Other explanations}

A rather simple alternative explanation is that the low UV scattering
efficiency and gray scattering described above could be the intrinsic
properties of the dust grain population in the circumnuclear region of
active galactic nuclei (AGNs).  Anomalous properties of dust grains in
the AGN vicinity have been reported by several authors, and these
properties have been summarized recently by Maiolino et al. (2001a).
They have shown that the dust grains in the circumnuclear region of 16
AGNs have a lower reddening $E_{B-V}$ than the standard Galactic dust
has for the column density $N_{\rm H}$ obtained from the X-ray
photoelectric absorption.  They also pointed out that in many cases
the optical absorption $A_V$ is also lower than expected from $N_{\rm
H}$, compared with the Galactic case.  The lower $E_{B-V} / N_{\rm H}$
leads to relatively gray scattering, and lower $A_V / N_{\rm H}$ leads
to lower UV/optical scattering efficiency.

These low $A_V / N_{\rm H}$ and $E_{B-V} / N_{\rm H}$ can be explained
by a dust size distribution dominated by large grains (Laor \& Draine
1993; Maiolino et al. 2001b).  This anomalous distribution could be
caused by the formation of large grains through coagulation in a
high-density environment (Maiolino et al. 2001b), and/or single-photon
destruction of small grains in the strong high-energy radiation field
(Laor \& Draine 1993). In the extended scattering region seen in our
images, the latter would be conceivable, though the former might not
be the case. It is possible that X-ray charging and heating of dust
grains by the AGN continuum is able to selectively destroy parts of
the dust population (Krolik \& Rhoads 2001).

\section{Conclusions}\label{sec-conc}

We have presented HST imaging polarimetry data of the Seyfert 2 galaxy
Mrk 3. The UV and near-UV radiation is highly polarized, and the
polarization position angle distribution is centrosymmetric,
consistent with scattering of radiation from a compact source.  We
have determined the location of the hidden nucleus as the center of
this centrosymmetric pattern. The polarized flux is extended out to
over $3''$ ($\sim 1$ kpc) from the nucleus, but most of it is from
the region within $\sim 1''$ ($\sim 300$ pc) from the nucleus.

The polarized flux distributions in the two broad band filters are
found to be different, and from these two images we have obtained the
color map of the polarized flux.  In the southern edge region, the
color is significantly redder than a Seyfert 1 color, indicating
reddening. In contrast, the color seems to be bluened in some regions,
suggesting optically-thin dust scattering.  However, some bright knots
have a color similar to that of a Seyfert 1 nucleus, which would imply
gray scattering.  The recent Chandra observation suggests rather large
amount of extended X-ray scattered continuum.  The ratio of the UV
scattered flux to this X-ray scattered flux is similar to that of
Seyfert 1s. This indicates rather low scattering efficiency in the UV,
much lower than the optically-thin scattering by Galactic dust. These
two properties, namely the gray scattering and low UV scattering
efficiency, might be explained by clumpy opaque dust scattering, or
alternatively these could be intrinsic to nuclear dust grains which
might have a size distribution dominated by large grains.  However, a
simple explanation by electron scattering is also viable.


\acknowledgments

Support for this work was provided by NASA through grant GO-6702 to L.
Kay from the Space Telescope Science Institute, which is operated by
AURA, Inc., under NASA contract NAS5-26555.  This work is based on
observations with the NASA/ESA Hubble Space Telescope, obtained at the
Space Telescope Science Institute.  The authors appreciate Masao Sako
who kindly looked at the Chandra data upon our request. The authors
also thank the referee for carefully reading the manuscript and
providing helpful comments.  This research has made use of the
NASA/IPAC Extragalactic Database (NED) which is operated by the Jet
Propulsion Laboratory, California Institute of Technology, under
contract with the National Aeronautics and Space
Administration. M.K. was a Guest User, Canadian Astronomy Data Centre,
which is operated by the Herzberg Institute of Astrophysics, National
Research Council of Canada.



\clearpage

\clearpage

\begin{deluxetable}{cclccrrrrrrr}
\tablecaption{Obtained FOC Data \label{tab-data}}
\tablewidth{0pt}
\tablehead{
\colhead{Rootname} & \colhead{Obs Date} &
\colhead{Filter} & \colhead{Exp Time (sec)} 
}
\startdata

x3md0101r & Dec 10, 1998 & F275W+POL0   & $5186.50$&\\
x3md0102r & Dec 10, 1998 & F275W+POL60  & $5186.50$&\\
x3md0103r & Dec 10, 1998 & F275W+POL120 & $5186.50$&\\
x3md0104r & Dec 10, 1998 & F342W+POL120 & $1696.50$&\\
x3md0105r & Dec 10, 1998 & F342W+POL60  & $1696.50$&\\
x3md0106r & Dec 10, 1998 & F342W+POL0   & $1696.50$&\\

\enddata
\end{deluxetable}


\newpage

\begin{figure}
\plotone{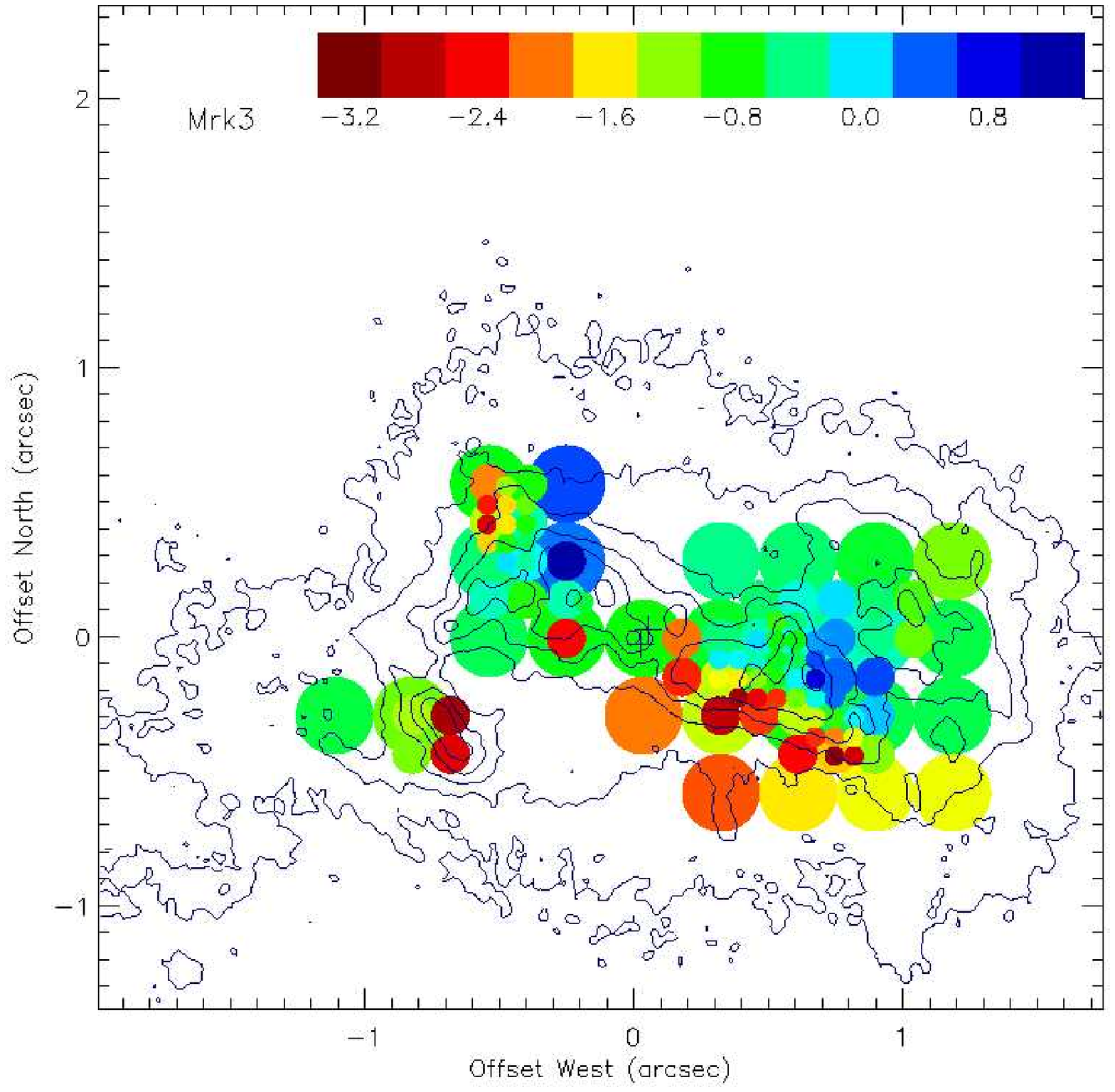}
\figcaption{Color version of Fig.\ref{fig_mrk3_pfcolor}}
\end{figure}

\begin{figure}
\plotone{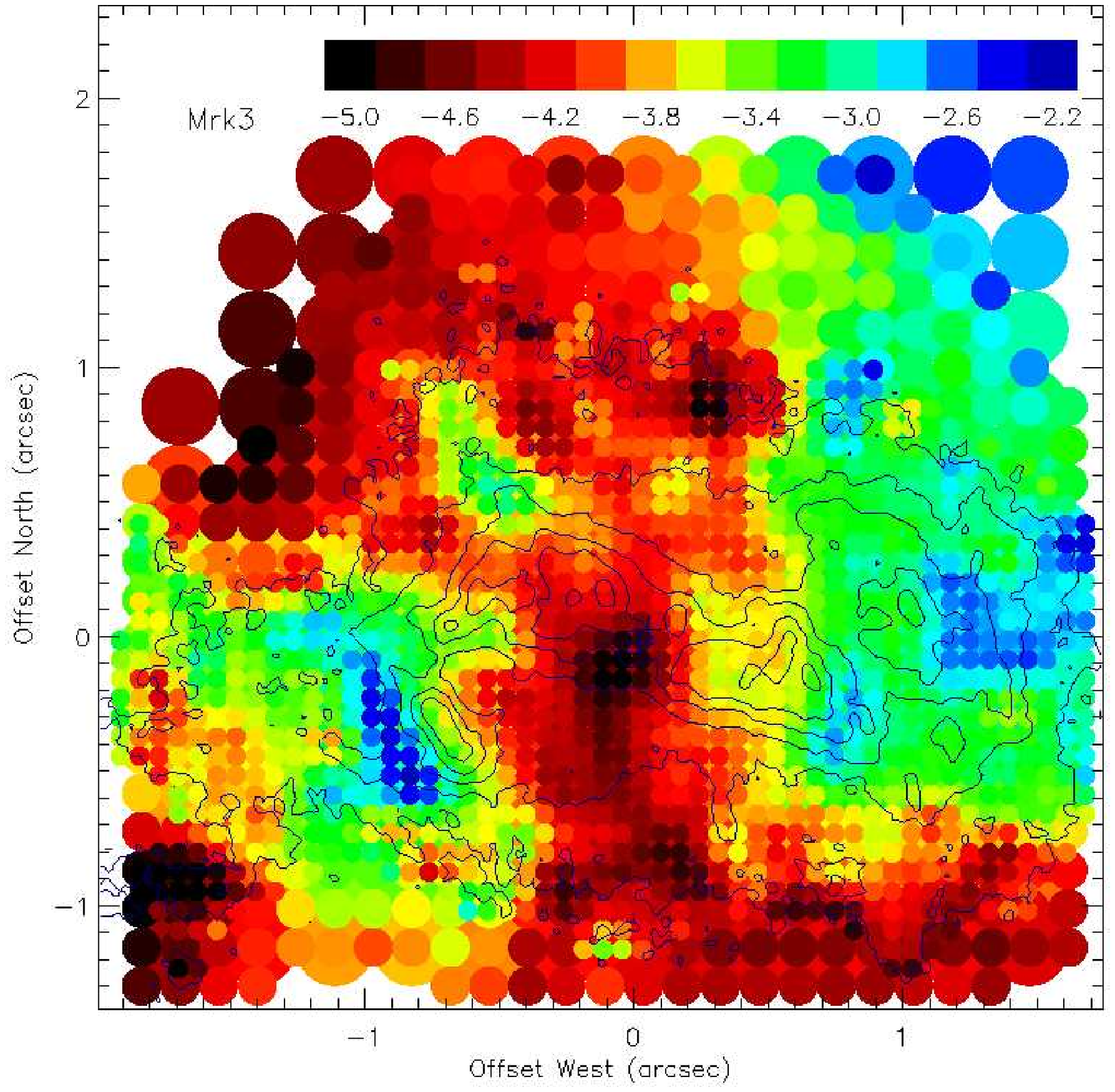}
\figcaption{Color version of Fig.\ref{fig_mrk3_tfcolor}}
\end{figure}

\end{document}